\documentstyle{article}[12pt]

\begin{document}

\baselineskip 12pt

\begin{flushleft}
{\Large \bf  Massive Gauge Fields and the Planck Scale}
\end{flushleft}
\vspace{0.5cm}
\begin{flushleft}
{\large \bf G.\ D.\ Acosta$^1$, M. Kirchbach$^2$}
\end{flushleft}
\vspace{0.05cm}

\begin{flushleft}
$^1${\it Instituto de Fis{\' {\i}}ca, UGTO,\\
Lomas del Bosque 103,
Fracc. Lomas de Campestre,\\
37150 Leon, Guanajuato, M\'exico}
\end{flushleft}

\begin{flushleft}
$^2${\it Instituto de Fis{\' {\i}}ca, UASLP,
Av. Manuel Nava 6,\\
Zona Universitaria,
San Luis Potos{\'{\i}},\\
SLP 78290, M\'exico}
\end{flushleft}
\begin{flushleft}
{ \it mariana@ifisica.uaslp.mx}\, (corr. auth.)\\
\end{flushleft}
\begin{flushleft}

\begin{flushleft}
The present work is devoted to massive gauge fields 
in special relativity with two fundamental constants- 
the velocity of light, and the Planck length, so called
doubly special relativity (DSR).
The two invariant scales are accounted for by properly modified
boost parameters. 
Within above framework we construct the vector potential as 
the $(1/2,0)\otimes (0,1/2)$ direct product, 
build the associated field strength tensor together with 
the Dirac spinors and use them to calculate  
various observables as functions of the Planck length.
\end{flushleft}

\vspace{0.05cm}
\begin{flushleft}
\end{flushleft}

{\bf 1. \hspace{0.1cm}INTRODUCTION}
\end{flushleft}
In the standard theory of special relativity,  Lorentz transformations
preserve the energy-momentum dispersion relation
of a particle observed from  two different inertial frames 
according to 
\begin{equation}
\frac{E}{c}^2- p_x^2- p_y^2- p_z^2=\left(\frac{E^\prime}{c} \right)^2- 
p_x^\prime \, ^2- p_y^\prime \, ^2- p_z^\prime\, ^2 =m^2c^2.\nonumber
\end{equation}
Lorentz transformations are covered by rotations in the three
space-like planes $(p_xp_y)$, $(p_yp_z)$, and $(p_zp_x)$, on the one side,
and by pseudo-rotations (i.e. rotations by an imaginary angle) in the 
$(\frac{E}{c},p_ x)$, $(\frac{E}{c},p_y)$, and 
$(\frac{E}{c}, p_z)$ planes, on the other side.
To be specific, for the text-book example of $p_y^\prime =p_y$, and 
$p_z^\prime =p_z$, the boost parametrizes as 
\begin{eqnarray}
\left(
\begin{array}{c}
\frac{E}{c}^\prime\\
p_x^\prime \\
p_y^\prime\\
p_z^\prime
\end{array}
\right)
=
\left(
\begin{array}{cccc}
\cosh \phi & \sinh \phi&0&0\\
\sinh\phi &\cosh\phi&0&0\\
0&0&1&0\\
0&0&0&1
\end{array}
\right)
\left(\begin{array}{c}
\frac{E}{c}\\
p_x\\
p_y\\
p_z
\end{array}
\right),\nonumber
\end{eqnarray}
\begin{equation}
\cosh\phi = \frac{E}{mc^2}\, , 
\quad   \sinh\phi =\frac{|\vec{p\, }|}{mc}\, .
\label{boost_SR}
\end{equation}
Ordinary special relativity, to be referred to as SR in the following,
shows up here through the dependence of
the boost parameters on the velocity of light. Usually, 
velocities are given in units of $c$, and one sets $c=1$.

In recent years, certain phenomena of Ultra High
Energy Cosmic Rays (UHECR) seem to indicate that in the vicinity of
the Planck length, ordinary special relativity may need an extension 
that accounts for the constancy of the Planck scale.
To be more specific, in effect of collisions with the soft photons
from the cosmic microwave background radiation, 
one expects cosmic ray protons, and cosmic gamma rays
to slow down to energies below $E_p<5\times 10^{19}$eV, and  
$E_\gamma <20$ TeV, respectively.
The UHECR protons slow down basically because
of pion-photo production, while the ultra high energy gamma rays
loose energy  due to electron-positron pair production.
Yet, cosmic protons with energies 
$E_p>  5\times 10^{19}$ eV (so called Greisen-Zatsepin-Kusmin (GZK) 
threshold value) as well as cosmic gamma rays with $E_\gamma> 20$TeV
still arrive at earth, and the GZK limit seems too low in
comparison to data \cite{9410067}.

A possible solution to this so called {\it cosmic ray problem\/}
has been advocated  in Refs.~ \cite{Amelino}, \cite{fb15}.
According to Amelino-Camelia, quantum gravity effects 
may force  deformations upon the energy-momentum dispersion relation 
as
\begin{equation}
E^2 \approx  c^2p^2 +c^4m^2 +\lambda _PEp^2 +
{\mathcal O}\left(\lambda_P^2 \right)\, .
\label{e_d_mod}
\end{equation}
Modifications of this type can allow for a higher value of the 
GZK limit \cite{Giov_Piran}, \cite{Kifune}, \cite{Ng_vanDam}. 
Special relativities with two invariant scales
are known as Doubly Special Relativity (DSR), a notion 
due to Ref.~ \cite{Amelino}.
The major idea of such theories of space-time is 
to replace the linear parametrization of the boost by a non-linear 
function of the Planck length without changing 
the algebra of the Lorentz group \cite{JV}. 
Amelino-Camelia's deformed energy-momentum dispersion relation 
results from  the following non-linear boost parametrization
(to be referred to as DSRa in the following)
\begin {equation}
\cosh \xi=\frac{e^{\lambda_{P} E}-\cosh(\lambda_{P} m)}
{\sinh (\lambda_{P} m)},\hskip 0.3in \sinh \xi=
\frac{\lambda_{P}|\vec p\, |~e^{\lambda_{P} E}}{\sinh (\lambda_{P} m)}.
\label{boost_AC}
\end{equation}
The energy momentum dispersion relation following from 
Eq.~(\ref{boost_AC}) reads
\begin{equation}
2\cosh\lambda_P E -\lambda_P^2\vec p \, ^2 e^{\lambda _PE}
=2\cosh \lambda_P m\, .
\label{E_p_AC}
\end{equation}
In a similar spirit, Smolin and Magueijo
proposed in Ref.~\cite{fb13} a different boost parametrization 
(to be referred to as DSRb) as 
\begin {equation}
\cosh \xi=\frac{E}{m}\frac{(1-\lambda_{P}m)}{(1-\lambda_{P}E)},\hskip 
0.3in \sinh \xi=\frac{\vert \vec p\, \vert }{m}
\frac{(1-\lambda_{P}m)}{(1-\lambda_{P}E)}\, .
\label{boost_SM}
\end{equation}
Both parameterizations  reduce to Eq.~(\ref{boost_SR}) at energies
significantly low compared to the Planck scale.
It is important to notice that the $\lambda_P$ parameter is real and 
positive \cite{Giovanni_lP}. {}For a broader
discussion of various DSR aspects the interested reader may wish
to consult Refs.~\cite{0203040}-\cite{0308028}.

The goal of the present paper is to obtain massive
Abelian gauge fields in special relativity with two 
invariant scales. The results easily extend to the
non-Abelian case. Massive gauge bosons are especially 
interesting in all field theories with spontaneously broken local
gauge symmetries such as the electroweak theory.
Moreover, they are the basic building blocks of massive gravitinos,
spin-3/2 gauge fermions that appear in supersymmetric theories
and which can have a significant impact onto models of
inflationary universe \cite{Nilles},
\cite{Kallosch}, and dark matter \cite{Bolz}.

The presentation is organized as follows.
In the next Section we discuss differences between 
special and doubly special relativities.
In Section 3 we build up the massive gauge field in DSR.
In Section 4 we construct the associated field strength tensor and
calculate various observables as a function of $\lambda_P$.
The paper closes with a brief summary and perspectives.

\begin{flushleft}
{\bf 2.\hspace{0.1cm} CPT VIOLATION IN DSR}
\end{flushleft}
The DSRb energy-momentum dispersion relation resulting
from Eq.~(\ref{boost_SM}) can be cast
into the form
\begin{equation}
E^2 \frac{(1-\lambda_P m)^2}{(1-\lambda_P E)^2 } -\vec p\, ^2
\frac{(1-\lambda_P m)^2}{(1-\lambda_PE)^2 }=m^2\, .
\label{E_p_SM}
\end{equation}
The main message from Eq.~(\ref{E_p_SM}) is the energy cut off
at $E=1/\lambda_P$.
Above this value (and for $m\not=1/\lambda_P$) 
 Eq.~(\ref{E_p_SM}) is equivalent  to
\begin{equation}
E^2-\vec p\, ^2 =  f^2(E) m ^2\, ,
\quad
f(E)= \frac{1-\lambda_PE}{1-\lambda_Pm }\, .
\label{E_p_SM2}
\end{equation}
In this way, the Planck length shows up only in the
energy dependent ``mass'' form factor.

Equation~(\ref{E_p_SM2}) describes two hyperboloids in the $(E,\vec p\, )$
space that differ from the ordinary SR hyperboloids in two aspects.
{}First, $(E^2-\vec p\, ^2)$ is no longer a constant, and second,
at rest, instead of the symmetric SR result, $E=\pm m$,
one finds the two asymmetric solutions 
\begin{equation}
E_1=m\, ,\qquad E_2=-\frac{m}{1-2\lambda_Pm }\, ,
\label{DSR_past_future}
\end{equation} 
respectively. The consequence of Eq.~(\ref{DSR_past_future}) is
CPT violation in Doubly Special Relativity,
an effect that reflects the non-locality introduced by the energy 
cut off. In DSRa, where the presence of $\lambda_P$ in the boost 
parameters does not result in discretization of the
phase space, one finds the symmetric solution
$E=\pm m$ at $\vec p\, =0$.

At this place a comment onto distinguishability between special-
and doubly special relativity seems to be in order.
{}From the formal point of view, one may redefine $ \cosh \xi $
and $\sinh \xi $ as \cite{JV}
\begin{eqnarray}
\cosh \xi =\frac{\epsilon}{\mu},
&\quad &
\sinh \xi =\frac{|\vec{\pi }|}{\mu}\, ,
\end{eqnarray}
and consider the quantities $\epsilon $, $\vec \pi $, and $\mu $
as the physical observables. As long as these new kinematic quantities
satisfy the $\epsilon^2 - \vec{\pi}^2=\mu^2  $ dispersion relation
of special relativity, one may feel tempted to claim 
``indistiguishability'' between SR and DSR.
Notice, however, that the $(\epsilon, \vec{ \pi})$ space
is CPT conserving because if rest is defined as
$\vec{ \pi} =0$ one finds the symmetric
$\epsilon=\pm \mu $ solution. The reason for this seeming contradiction
is that one forgets that contrary to $E$ and $\vec{p}$ in SR,
$\epsilon$ and $\vec{\pi }$ are discontinuous variables.
In effect, CPT distinguishes between SR and DSR.

It would be interesting to check consistency of
Eq.~(\ref{DSR_past_future}) with the 
measured proton-anti-proton mass difference for which the
The Particle Data Group reports in Ref.~ \cite {PART} 
the following value:
\begin{equation}
\frac{|m_p-m_{\overline{ p}}|}{m_p} <  6\times 10^{-8}\, .
\label{CPT_p_ap}
\end{equation}
Incorporating  Eq.~(\ref{DSR_past_future}) into ~(\ref{CPT_p_ap})
and using  $m_p=938.271998$ MeV for the proton mass amounts to
\begin{equation}
\lambda_P < 0.6309 \times 10^{-23} \mbox{m}\, .
\label{Vergleich}
\end{equation}
This inequality     is clearly satisfied by the theoretical 
value of $\lambda_P\sim 10^{-35}$ m.

\begin{flushleft}
{\bf 3.\hspace{0.1cm} THE VECTOR POTENTIAL AT THE PLANCK SCALE}
\end{flushleft}
\begin{flushleft}
{\bf 3.1\hspace{0.1cm}Spinor and Co-Spinor  Representations
at the Planck scale}
\end{flushleft} 
As long as one maintains the algebra of the Lorentz group intact,
the group representations remain unaltered. 
We here are especially interested in the spinor $(1/2,0)$--, and co-spinor--,
$(0,1/2)$, representations, respectively. 
In what follows we shall use the notation
$\zeta_h (\vec{p\, })$ and ${\dot \zeta}_h (\vec p\, )$ 
(with $h$ standing for spin-projection ) for
spinor $(1/2,0)$, and co-spinor $(0,1/2)$, respectively.
In the Cartesian frame,
\begin{eqnarray}
\zeta_\uparrow (\vec{0\, })=\sqrt{m}
\left(
\begin{array}{c}
1\\
0
\end{array}
\right)\, ,
&\qquad &
\zeta_\downarrow (\vec{0\, })=\sqrt{m}
\left(
\begin{array}{c}
0\\
1
\end{array}
\right)\, ,\nonumber\\
\dot \zeta _\uparrow (\vec{0\,})= \zeta _\uparrow (\vec{0\,})\, ,
&\qquad &
\dot \zeta _\downarrow (\vec{0\,})= \zeta _\downarrow (\vec{0\,})\, .
\label{Cart_spinors}
\end{eqnarray}

According to standard rules \cite{Hladek} spinors and
co-spinors are boosted as
\begin{eqnarray} \label{E:estimd} 
\zeta_h(\vec{p\, })&=&\exp \left(+\frac{\vec{\sigma}}{2} \cdot 
\vec{\xi}\right)
\zeta_h(\vec{0\, }),
\\ \label{E:estimi}
{\dot \zeta }_h(\vec{p\, })&=&
\exp \left(-\frac{\vec{\sigma}}{2} \cdot \vec{\xi}\right)
{\dot \zeta }_h(\vec{0}),\quad h=\uparrow\, \downarrow\, ,
\end{eqnarray}
where $\sigma _i$, $i=1,2,3$,  are the standard Pauli matrices.

In writing down the exponentials in Eq.~(\ref{E:estimd})
and (\ref{E:estimi}) as
\begin{eqnarray}
\exp(\pm \frac{\vec{\sigma}}{2}\cdot\vec{\xi})&=&
I_{2}\cosh \frac{\xi}{2} \pm(\vec{\sigma}\cdot\hat{\xi})\sinh \frac{\xi}{2},
\end{eqnarray}
one finds
\begin{eqnarray}
\exp(\pm \frac{\vec{\sigma}}{2}\cdot\vec{\xi})&=&
l[I_{2}\textup{B} \pm (\vec{\sigma}\cdot\hat{\xi})\Gamma]\, .
\end{eqnarray}
Here, $l=(\textup{B}^{2}-\Gamma^{2})^{-1/2}$. The previous equation is
of great importance for the present work, as the new observer independent 
scale, the Planck length, appears for the first time in the building blocks
of the vector potential via 
\begin{eqnarray}
\textup{B}^2=e^{\lambda_{P} E}\!\! -e^{-\lambda_{P} m}, &\qquad&
\Gamma^2=e^{\lambda_{P} E}-e^{\lambda_{P} m}\, \quad 
\mbox{in DRSa}, \\
\textup{B}^2=E+m-2mE \lambda_{P}, &\qquad& \Gamma^2=E-m\, .
\quad\qquad\,\mbox{in DSRb}.
\label{B_G_DSR}
\end{eqnarray} 

\begin{flushleft}
{\bf 3.2\hspace{0.1cm}The Vector Potential as $(1/2,0)\otimes (0,1/2)$ }
\end{flushleft}
At rest, and in the Cartesian frame,
the basis vectors of the product space are given by
\begin{eqnarray}
a_{\uparrow\uparrow}(\vec{0\, })=m
\left(\begin{array}{c}
1\\
0
\end{array}\right)
\otimes 
\left(\begin{array}{c}
1\\
0
\end{array}\right)\, ,
&\quad &
a_{\downarrow\downarrow}(\vec{0\, })
=m
\left(\begin{array}{c}
0\\
1
\end{array}\right)
\otimes 
\left(\begin{array}{c}
0\\
1
\end{array}\right)\, ,\nonumber\\
a_{\uparrow\downarrow }(\vec{0\, })= m
\left(\begin{array}{c}
1\\
0
\end{array}\right)
\otimes 
\left(\begin{array}{c}
0\\
1
\end{array}\right)\, ,
&\quad &
a_{\downarrow\uparrow }(\vec{0\, })= m
\left(\begin{array}{c}
0\\
1
\end{array}\right)
\otimes 
\left(\begin{array}{c}
1\\
0
\end{array}\right)\, .
\label{primt_bas}
\end{eqnarray}
In order to obtain rest frame basis of good spin, one needs
to subject the direct-product basis to the
following transformation: 
\begin{eqnarray}
\epsilon_{JM}(\vec 0\, )={\mathcal S}S_1 a_{h, h^\prime }(\vec 0\, )\, ,
\quad S_1&=&
\left(
\begin{array}{cccc}
1&0&0&0\\
0&{1\over \sqrt{2}}&{1\over \sqrt{2}}&0\\
0&-{1\over \sqrt{2}}&{1\over \sqrt{2}}&0\\
0&0&0&1
\end{array}
\right)\, , \quad
{\mathcal S}=\frac{1}{\sqrt{2}}\left(
\begin{array}{cccc}
0&1&-1&0\\
-1&0&0&1\\
-i&0&0&-i\\
0&1&1&0
\end{array}
\right)\, .\nonumber\\
\end{eqnarray} 
Here, 
$\epsilon_{0,0}(\vec 0\, )={\mathcal S} S_1a_{\uparrow\downarrow}(\vec 0\, )$, 
is a scalar, while 
$\epsilon_{1, 0}(\vec 0\, )={\mathcal S}S_1a_{\downarrow\uparrow}(\vec 0\, )$,
 $\epsilon_{1, 1}(\vec 0\, )={\mathcal S}S_1 a_{\uparrow\uparrow}(\vec 0\, )$, 
and 
$\epsilon_{1, -1}(\vec 0\, )={\mathcal S}S_1 a_{\downarrow\downarrow}
(\vec 0\, )$
constitute the spin-triplet.
The latter three vectors describe in turn
longitudinal-- , right-- , and left-handed circularly polarized gauge bosons. 

The boost in the product space is obtained
as the direct product of the spinor- and co-spinor 
boosts and reads:
\begin{equation}
\kappa (\vec {p}\, )=\exp\left(+\frac{1}{2}\vec{\sigma}\cdot 
\vec {\xi} \right)\otimes\exp\left(-\frac{1}{2}\vec{\sigma}\cdot
\vec{\xi}\right).
\end{equation}
Using the DSR boost parameterizations from the previous subsection
amounts to
\begin{eqnarray}
\kappa (\vec{p}\, )=l^{2}\left(\matrix{
\textup{B}^{2}-\Gamma^{2}\hat{p}^{2}_{z}  & -\Gamma\hat{p}_{-}
(\textup{B}+\Gamma \hat{p}_{z}\, ) 
& \Gamma\hat{p}_{-}(\textup{B}-\Gamma \hat{p}_{z}\, )
& -\Gamma^{2}\hat{p}_{-}^{2}\cr 
-\Gamma\hat{p}_{+}(\textup{B}+\Gamma \hat{p}_{z}\, )  & 
(\textup{B}+\Gamma \hat{p}_{z}\, )^{2}
 & -\Gamma^{2}(\hat{p}_{x}^{2} + \hat{p}_{y}^{2}\, ) & 
\Gamma\hat{p}_{-}(\textup{B}+\Gamma \hat{p}_{z}\, )\cr
\Gamma\hat{p}_{+}(\textup{B}-\Gamma \hat{p}_{z}\, ) & -
\Gamma^{2}(\hat{p}_{x}^{2} + \hat{p}_{y}^{2})
 & (\textup{B}-\Gamma \hat{p}_{z})^{2} & \Gamma\hat{p}_{-}
(-\textup{B}+\Gamma \hat{p}_{z})\cr
-\Gamma^{2}\hat{p}_{+}^{2} & \Gamma\hat{p}_{+}(\textup{B}+\Gamma \hat{p}_{z})
 & \Gamma\hat{p}_{+}(-\textup{B}+\Gamma \hat{p}_{z}) & 
\textup{B}^{2}-\Gamma^{2}\hat{p}^{2}_{z}
\cr} \right) ,\nonumber\\
\end{eqnarray}
where, $\hat{p}_{\pm}=\hat{p}_{x} \pm i\hat{p}_{y}$, and 
$\hat{p}_{\zeta}=p_{\zeta}/\vert \vec p\, \vert $. 
Notice that same calculus has been exploited in Ref.~\cite{0101009}
to construct a massive electromagnetic  field in ordinary special
relativity.
The operator that boosts $\epsilon_{JM}(\vec 0\, )$ is
\begin{equation}
\kappa^\prime (\vec p\, )=
{\mathcal S}S_1 \kappa (\vec p\, )\left(S_1{\mathcal S}\right)^{-1}\, .
\end{equation} 

As long as neither $J^2$, nor $J_3$ commute with
$\kappa^\prime  (\vec{p\, })$,  the boosted vectors 
$\kappa^\prime  (\vec{p}\, )\epsilon_{JM}(\vec 0\, )$  
can not any longer be characterized by the $J$, and $M$ 
quantum numbers but have to be labeled differently.
Below we introduce following notations:
\begin{eqnarray}
{\mathcal A}_{1}^{\mu}(\vec{p}\, ) & := & \kappa^\prime(\vec p\, )
\epsilon_{1,1}(\vec 0)=
-i \frac{l^{2}m}{\sqrt{2}}\left(\matrix{
2\textup{B}\Gamma(-i\hat{p}_{x}+\hat{p}_y) \cr
-i(\textup{B}^{2}+\Gamma^{2}[(\hat{p}_{x}+
i\hat{p}_{y})^{2}-\hat{p}_{z}^{2}]) \cr
\textup{B}^{2}+\Gamma^{2}[-(\hat{p}_{x}+i\hat{p}_{y})^{2}-\hat{p}_{z}^{2}] \cr
2\Gamma^{2}(-i\hat{p}_{x}+\hat{p}_y)\hat{p}_{z}
}\right),\\
{\mathcal A}_{2}^{\mu}(\vec{p}\, ) & := &
\kappa^\prime(\vec p\, )\epsilon_{1,0}(\vec 0)= ml^{2}\left(\matrix{
2\textup{B}\Gamma\hat{p}_{z} \cr
2\Gamma^{2}\hat{p}_{x}\hat{p}_{z} \cr
2\Gamma^{2}\hat{p}_{y}\hat{p}_{z} \cr
\textup{B}^{2}-\Gamma^{2}(\hat{p}_{x}^{2}+\hat{p}_{y}^{2}-\hat{p}_{z}^{2})
}\right),\\
{\mathcal A}_{3}^{\mu}(\vec{p}\, ) & := & 
\kappa^\prime(\vec p\, )\epsilon_{1,-1}(\vec 0)=
-i\frac{l^{2}m}{\sqrt{2}}\left(\matrix{
2\textup{B}\Gamma(i\hat{p}_{x}+\hat{p}_y) \cr
i(\textup{B}^{2}+\Gamma^{2}[(\hat{p}_{x}-
i\hat{p}_{y})^{2}-\hat{p}_{z}^{2}]) \cr
\textup{B}^{2}+\Gamma^{2}[-(\hat{p}_{x}-i\hat{p}_{y})^{2}-\hat{p}_{z}^{2}] \cr
\Gamma^{2}(i\hat{p}_{x}+\hat{p}_y)\hat{p}_{z} 
}\right),\\
{\mathcal A}_{4}^{\mu}(\vec{p}\, ) & := & 
\kappa^\prime(\vec p\, )\epsilon_{0,0}(\vec 0)= 
ml^{2}\left(\matrix{
\textup{B}^{2}+\Gamma^{2} \cr
2\textup{B}\Gamma\hat{p}_{x} \cr
2\textup{B}\Gamma\hat{p}_{y} \cr
2\textup{B}\Gamma\hat{p}_{z}
}\right).
\end{eqnarray}
The massive gauge vector-potential is now completely described
by the last expressions, which are the first result
of the  present study. Knowing the 
basis in $(1/2,0)\otimes (0,1/2)$ in DSR allows for the construction
of the DSR field strength tensor, a topic considered in the next
Section.

\begin{flushleft}
{\bf 4.\hspace{0.1cm} FIELD  STRENGTH TENSOR IN DSR}
\end{flushleft}
The field strength tensor, ${\mathcal F}^{\mu \nu} 
(\vec{p\, }, \zeta )$,
(with $\zeta =0,1,2,3$)
is constructed from ${\mathcal A}_i(\vec{p\, })$ as a solution of the
massive Proca equation
\begin{eqnarray}\label{E:eq1}
p_{\mu}{\mathcal F}^{\mu \nu}(\vec{p},\zeta)=m^2e^{\nu}_{\zeta}(\vec {p}\, ).
\end{eqnarray}
Here, we defined
\begin{eqnarray} \label{E:eq2}
&e_{0}(\vec{p}\, )= {\mathcal A}_{4}(\vec{p}\, ),  &
e_{1}(\vec{p}\, )=\frac{1}{\sqrt{2}}\left({\mathcal A}_{3}(\vec{p}\, )-
{\mathcal A}_{1}(\vec{p}\, )\right), 
\nonumber \\ &e_{3}(\vec{p}\, )={\mathcal A}_{2}(\vec{p}\, ), &\,\,
e_{2}(\vec{p}\, )=\frac{i}{\sqrt{2}}\left({\mathcal A}_{3}(\vec{p}\, )+
{\mathcal A}_{1}(\vec{p}\, )\right).
\end{eqnarray}
Equation.~(\ref{E:eq1}) translates into a matrix form as 
\begin{eqnarray}
{\mathcal F}(\vec{p},\zeta)\, G\, e_{0}(\vec{p}\, )=m^2 
e_{\zeta}(\vec{p}\, ),
\label{FT_MTR}
\end{eqnarray}
where the $G$ matrix stands for the metric tensor
$G=$diag$(1,-1,-1,-1)$.
As a  solution of Eq.~(\ref{FT_MTR}) one finds  (see Refs.~\cite{Valery}, 
\cite{Bety} for more details)
\begin{eqnarray}
{\mathcal F}(\vec{p},\zeta)=e_{\zeta}(\vec{p}\, )e_{0}^{\dagger}(\vec{p}\, )-
e_{0}(\vec{p}\, )e_{\zeta}^{\dagger}(\vec{p}\, ).
\label{stress_t}
\end{eqnarray}
In DSR, the explicit expressions for $e_\zeta (\vec{p\, })$ are given by
\begin{eqnarray} \label{E:eq2a}
&e_{1}(\vec{p}\, ) =ml^{2}\left(\matrix{
2\textup{B}\Gamma \hat{p}_{x} \cr
\textup{B}^{2}-\Gamma^{2}(1-2\hat{p}_{x}^{2})\cr
2\Gamma^{2} \hat{p}_{x}\hat{p}_{y} \cr
2\Gamma^{2} \hat{p}_{x}\hat{p}_{z}
}\right), 
&e_{2}(\vec{p}\, )=ml^{2}\left(\matrix{
2\textup{B}\Gamma \hat{p}_{y} \cr
2\Gamma^{2} \hat{p}_{x}\hat{p}_{y} \cr
\textup{B}^{2}-\Gamma^{2}(1-2\hat{p}_{y}^{2})\cr
2\Gamma^{2} \hat{p}_{y}\hat{p}_{z}
}\right), \nonumber \\
&e_{3}(\vec{p}\, )=ml^{2}\left(\matrix{
2\textup{B}\Gamma \hat{p}_{z} \cr
2\Gamma^{2} \hat{p}_{x}\hat{p}_{z} \cr
2\Gamma^{2} \hat{p}_{y}\hat{p}_{z} \cr
\textup{B}^{2}-\Gamma^{2}(1-2\hat{p}_{z}^{2})
}\right),
&e_{0}(\vec{p}\, ) =ml^{2}\left(\matrix{
\textup{B}^{2}+\Gamma^{2} \cr
2\textup{B}\Gamma\hat{p}_{x} \cr
2\textup{B}\Gamma\hat{p}_{y} \cr
2\textup{B}\Gamma\hat{p}_{z}
}\right)\, ,
\end{eqnarray}
with $l^2=\textup{B}^{2}-\Gamma^{2}$  from Eq.~(\ref{B_G_DSR}).
Equation~(\ref{stress_t}) allows to construct
the massive field strength tensor in momentum space.
In so doing one finds:
\begin{eqnarray} 
{\mathcal F}(\vec{p},x)  =  m^2l^{2}\left(\matrix{
0
&-(\textup{B}^{2}+\Gamma^{2})+2\Gamma^{2}\hat{p}_{x}^{2}
&2\Gamma^{2}\hat{p}_{x}\hat{p}_{y}
&2\Gamma^{2}\hat{p}_{x}\hat{p}_{z} \cr
(\textup{B}^{2}+\Gamma^{2})-2\Gamma^{2}\hat{p}_{x}^{2}
&0
&2\textup{B}\Gamma\hat{p}_{y}
&2\textup{B}\Gamma\hat{p}_{z} \cr
-2\Gamma^{2}\hat{p}_{x}\hat{p}_{y}
&-2\textup{B}\Gamma\hat{p}_{y}
&0
&0 \cr
-2\Gamma^{2}\hat{p}_{x}\hat{p}_{z}
&-2\textup{B}\Gamma\hat{p}_{z}
&0
&0
}\right), \nonumber
\end{eqnarray} 
\begin{eqnarray} 
{\mathcal F}(\vec{p},y)  =  m^{2}l^{2}\left(\matrix{
0
&2\Gamma^{2}\hat {p}_{x}\hat{p}_{y}
&-(\textup{B}^{2}+\Gamma^{2})+2\Gamma^{2}\hat{p}_{y}^{2}
&2\Gamma^{2}\hat{p}_{y}\hat{p}_{z} \cr
-2\Gamma^{2}\hat{p}_{x}\hat{p}_{y}
&0
&-2\textup{B}\Gamma \hat {p}_{x}
&0 \cr
(\textup{B}^{2}+\Gamma^{2})-2\Gamma^{2}\hat{p}_{y}^{2}
&2\textup{B}\Gamma\hat{p}_{x} 
&0
&2\textup{B}\Gamma\hat{p}_{z} \cr
-2\Gamma^{2}\hat{p}_{y}\hat{p}_{z}
&0
&-2\textup{B}\Gamma\hat{p}_{z}
&0
}\right), \nonumber
\label{stress_t_x}
\end{eqnarray}
\begin{eqnarray} \label{E:tensr2}
{\mathcal F}(\vec{p},z)  =  m^{2}l^{2}\left(\matrix{
0
&2\Gamma^{2}\hat{p}_{x}\hat{p}_{z}
&2\Gamma^{2}\hat{p}_{y}\hat{p}_{z} 
&-(\textup{B}^{2}+\Gamma^{2})+2\Gamma^{2}\hat{p}_{z}^{2} \cr
-2\Gamma^{2}\hat{p}_{x}\hat{p}_{z}
&0
&0 
&-2\textup{B}\Gamma\hat{p}_{x} \cr
-2\Gamma^{2}\hat{p}_{y}\hat{p}_{z}
&0
&0
&-2\textup{B}\Gamma\hat{p}_{y} \cr
(\textup{B}^{2}+\Gamma^{2})-2\Gamma^{2}\hat{p}_{z}^{2} 
&2\textup{B}\Gamma\hat{p}_{x}
&2\textup{B}\Gamma\hat{p}_{y}
&0
}\right)\, .
\nonumber\\
\end{eqnarray}
The correspondence between the field strength tensor components and
the vector  and axial vector fields, in turn denoted by
$\vec{\mathcal E\, }(\vec{p},\zeta )$, and 
$\vec{\mathcal B\, }(\vec{p},\zeta )$, is standard and reads
 \begin{eqnarray}
{\mathcal F}^{\mu \nu}(\vec{p}, \zeta )= \left( \matrix{
0                 &{\mathcal E}_{x}(\vec{p} ,\zeta ) &
{\mathcal E}_{y}(\vec{p},\zeta  )  
&{\mathcal E}_{z}(\vec{p},\zeta  )\cr
-{\mathcal E}_{x}(\vec{p},\zeta  )  &0              
&-{\mathcal B}_{z}(\vec{p},\zeta  ) 
&{\mathcal B}_{y}(\vec{p},\zeta )\cr
-{\mathcal E}_{y}(\vec{p},\zeta )  &{\mathcal B}_{z}(\vec{p},\zeta ) &0               
&-{\mathcal B}_{x}(\vec{p},\zeta )\cr
-{\mathcal E}_{z}(\vec{p},\zeta  )  &-{\mathcal B}_{y}(\vec{p},\zeta )&
{\mathcal B}_{x}(\vec{p},\zeta )  &0
}\right). 
\label{stress_E_B}
\end{eqnarray}
Equations~(\ref{E:tensr2}) represent a basis for
any massive Abelian field strength tensor near Planck energy.

\begin{flushleft}
{\bf 4.1 The $\lambda_P\to 0$ Limit of the Field Strength Tensor }
\end{flushleft}
Before proceeding further it is necessary to verify that 
Eqs.~(\ref{E:tensr2})--~(\ref{stress_E_B})
also qualify for the description of massless gauge fields in special
relativity, i.e. for  $\lambda_P\to 0$, and $m\to 0$  they have to
reduce to the standard electromagnetic field strength tensors.
In other words, one needs to confirm several established
relations of standard electrodynamics such as
covariant normalization of $F_{\mu\nu}(\vec p\, ,\zeta)$,
the divergences of the electric, and magnetic fields, and
the continuity conditions. In so doing we find
\begin{eqnarray}
-\frac{1}{ 4} 
F^{\mu\nu}(\vec p\, ,\zeta )
F_{\mu\nu}(\vec p\, ,\zeta ) &=&
\frac{1}{2}\left( 
\vec {\mathcal E}^2\,  (\vec p\, ,\zeta )-
\vec {\mathcal B}\,  ^2(\vec p\,  ,\zeta )\right) 
=\frac{E^2-\vec p\, ^2}{2 }\stackrel{{m\to 0}}{\longrightarrow} 0\,,
\nonumber\\
\label{cov_norm_SR}\\
\mbox{div} \vec {\mathcal B}\, (\vec p\, ,\zeta )=0\, ,
&\quad&
\mbox{div}\vec {\mathcal E}\, (\vec p\, ,\zeta )=-m^2p_\zeta\, 
\stackrel{m\to 0}{\longrightarrow} 0\, .
\end{eqnarray}
Notice that as long as our vector potentials are in the Lorentz gauge,
the massive electric field is not necessarily divergence-less.
Moreover, one can calculate the (classical) energy density of  
electromagnetic field, $U(\vec p\, ,\zeta )$,
together with the Poynting vector, $\vec{S}(\vec p\, ,\zeta )$,
and check the continuity condition,
\begin{eqnarray}
-U(\vec p\, , \zeta \, ) &=&
\frac{1}{2m}\left( \vec {\mathcal E}\, (\vec p\, ,\zeta  )^2 +
\vec {\mathcal B}\, (\vec p\, ,\zeta  )^2
\right)\, =\frac{1}{2m}\left(
2E^2 -m^2 - 2\frac{E^2-m^2}{p^2}p_\zeta ^2\right)\, ,
\nonumber\\
\vec S\, (\vec p\, ,\zeta  )&=&\frac{1}{m}
\vec {\mathcal E}\, (\vec p\, ,\zeta  )\times \vec {\mathcal B}\, 
(\vec p\, ,\zeta  )\, .
\label{CEQ_INPT}
\end{eqnarray}
By means of Eq.~(\ref{CEQ_INPT}) one calculates the continuity equation
as
\begin{equation}
E\, U(\vec p\, ,\zeta )- \vec p\, \cdot \vec S\, (\vec p\, ,\zeta )=
-\frac{E}{2} m \, 
\stackrel{m\to 0}{\longrightarrow} 0\, .
\label{cont}
\end{equation} 
Therefore, our construct for the  DSR field strength tensor 
reproduces correctly the massless electrodynamics limit. 

\begin{flushleft}
{\bf 4.2\hspace{0.1cm} Covariant Norm and 
Energy Density of Massive Gauge Fields in DSR}
\end{flushleft}
Equipped with the confidence of having created a construct
consistent with standard electrodynamics, we are now in a position
to calculate DSR corrections to the classical field energy density at
the Planck scale as:
 \begin{eqnarray} \label{ex}
\frac{1}{2m}\left( \vec{\mathcal E}\, ^2
(\vec{p},\zeta )+\vec{\mathcal B}\, ^2 (\vec{p},\zeta )
\right)=
\frac{\textup{B}^4+6
\textup{B}^2\Gamma^2+8\textup{B}^2\Gamma^2 \hat{p}_{\zeta}}
{2(\textup{B}^2-\Gamma^2)^2},
\end{eqnarray}
for each one of the tensors (\ref{E:tensr2}). 
This expression makes manifest  the modification of
the electromagnetic field energy density at the Planck scale.
To  first order on $\lambda_P$, Eq.~(\ref{ex}) takes the form 
\begin{eqnarray}\label{ap} 
\frac{1}{2m}\left( \vec{{\mathcal E}}\, ^2 (\vec{p},\zeta )
+\vec{{\mathcal B}}\, ^2(\vec{p},\zeta ) \right)=
\frac{1}{2m}\left(2E^2-m^2-2(E^2-m^2) \frac{ p_{\zeta}^2}{\vec p\, ^2} +
\delta_{\zeta}\lambda_P\right),
\label{U_DSR}
\end{eqnarray}
with:
\begin{eqnarray}
\delta_{\zeta}=\cases{{2E(E^2-m^2)(1-\frac{p_{\zeta}^2}{\vec p\, ^2})} 
&in DRSa, \cr 4E^2(E-m)(1- \frac{p_{\zeta}^2}{\vec p\, ^2}) &in DSRb.\cr}
\end{eqnarray}
In the expressions above $\zeta $ takes the values $\zeta= 1,2,3$.
Equation~(\ref{U_DSR}) is valid for low energy 
Abelian gauge fields with masses comparable to 
$\lambda_P^{-1}$ as they can appear in supersymmetric theories.

\noindent
The DSR counterpart of Eq.~(\ref{cov_norm_SR}) is
\begin{equation}
-\frac{1}{4}F^{\mu\nu}F_{\mu \nu }=\frac{1}{2}m^2\, .
\label{cov_norm_DSR}
\end{equation}
The normalization of the field strength tensor, 
in being brought about by the normalization of the 
(1/2,1/2) basis vectors \cite{Valery} in Eq.~(\ref{primt_bas})), 
comes out same in DSR and SR though in the former case 
we did not have to make use of the corresponding 
energy-momentum dispersion relation. 
That in the rhs in Eq.~(\ref{cov_norm_DSR}) one finds $m^2$,
the SR mass,  and not $\mu^2$, the DSR mass counterpart, 
is one more argument in favor of distinguishability
between SR and DSR.

\begin{flushleft}
{\bf 4.3\hspace{0,1cm}Energy Dependent Charge in DSR}
\end{flushleft} 
Another interesting observable is the vector Noether charge 
probed by the massive gauge field under consideration.
In order to calculate it, one first needs to construct
the DSR matter fields. We here construct the Dirac spinors
in DSR and calculate the time component of the vector current.
To do so we first notice equality of rest- frame spinor
and co-spinor 
$
\zeta_{\uparrow}(\vec{0}\, )= \zeta_{\downarrow}(\vec{0}\, )\, ,
$
and introduce the Dirac rest frame spinors, $u_h(\vec 0\, )$,
in the standard way as the direct sum of a 
spinor--and co-spinor spaces \cite{Hladek}:
\begin{eqnarray}
u_{\uparrow}(\vec{p}\, )=k(\vec{p}\, )\frac{1}{\sqrt{2}}
\left(\matrix{ 1 \cr 0 \cr 1 \cr 0}\right),
~~~~ u_{\downarrow}(\vec{p}\, )= k(\vec{p}\, )\frac{1}{\sqrt{2}}
\left(\matrix{ 0 \cr 1 \cr 0 \cr 1 }\right).
\end{eqnarray}
The boost in the direct sum space is standard and given by 
\begin{equation}
k (\vec {p}\, )=\exp\left(+\frac{1}{2}\vec{\sigma}\cdot 
\vec {\xi} \right)\oplus\exp\left(-\frac{1}{2}\vec{\sigma}
\cdot\vec{\xi}\right),
\end{equation}
with the result
\begin{eqnarray}
k (\vec{p}\, )=l\left(\matrix{
\textup{B}+\Gamma\hat{p}_{z}  & \Gamma(\hat{p}_{x}-i\hat{p}_y) 
&0 &0
\cr 
\Gamma(\hat{p}_{x}+i\hat{p}_y) &\textup{B}-\Gamma\hat{p}_{z}
 &  0  &0\cr
0 & 0
 &\textup{B}-\Gamma\hat{p}_{z} & -\Gamma(\hat{p}_{x}-i\hat{p}_y)\cr
0 & 0
 & -\Gamma(\hat{p}_{x}+i\hat{p}_y) & \textup{B}+\Gamma\hat{p}_{z}
\cr} \right)\, .\nonumber\\
\end{eqnarray}
In effect,  one obtains the following expressions for the boosted
Dirac particle spinors in DSR:
\begin{eqnarray}
u_{\uparrow}(\vec{p}\, )=\frac{l}{\sqrt{2}}
\left(\matrix{ \textup{B}+\Gamma\hat{p}_{z}\cr 
\Gamma(\hat{p}_{x}+i\hat{p}_y) \cr (\textup{B}-
\Gamma\hat{p}_{z}) \cr - 
\Gamma(\hat{p}_{x}+i\hat{p}_y)}\right),
~~ u_{\downarrow}(\vec{p}\, )=\frac{l}{\sqrt{2}}
\left(\matrix{ \Gamma(\hat{p}_{x}-i\hat{p}_y) \cr 
\textup{B}-\Gamma\hat{p}_{z} \cr -
\Gamma(\hat{p}_{x}-i\hat{p}_y) \cr 
(\textup{B}+\Gamma\hat{p}_{z})}\right)\, .
\label{Dir_sp_DSR}
\end{eqnarray}
In the $\lambda_P\to 0$ limit, Eqs.~(\ref{Dir_sp_DSR}) reduce to
ordinary Dirac spinors in special relativity. 
The antiparticle spinors, $v_h (\vec{p}\, )$ are then obtained as
$v_h(\vec{p}\, )=\gamma_5 u_h(\vec{p}\, )$, with
$\gamma_5$=diag$(1,1,-1,-1)$.

We now exploit Eq.~(\ref{Dir_sp_DSR}) to calculate the
vector charge density which at zero three momentum transfer is given by
\begin{equation}
Q(E )=\bar u_h(\vec p\, ) \gamma_0u_h(\vec p\, )\,=\cosh \zeta \, ,
\quad
\gamma_0=\left(\begin{array}{cc}
0_2&1_2\\
1_2&0_2
\end{array}
\right)\, . 
\end{equation}
{}For the  DSR/SR charge density ratio  one finds
\begin{equation}
\frac{Q_{DSR}(E \, )}{Q_{SR}(E \, )}
==\cases{{ \frac{1-\lambda_P m}{1-\lambda_P E}} 
&in DRSb, \cr \frac{m}{E}
\frac{e^{\lambda_PE}-\cosh(\lambda_P m)}{\sinh(\lambda_P m)} &in DSRa.\cr}
\label{charge_screening}
\end{equation}
In other words, compared to SR,
the DSR Noether charge density appears anti-screened in DSRb.

\begin{flushleft}
{\bf 5.\hspace{0.1cm}SUMMARY AND PERSPECTIVES}
\end{flushleft}
In this work we elaborated matter--, and massive gauge fields
in a space-time with two fundamental scales- the velocity of 
light, and the Planck length. Thereby we provide the basics of
a quantum field theory within the scenario of 
Doubly Special Relativity.
Once having Dirac and  gauge spinors at our disposal,
various boson-fermion couplings like
\begin{eqnarray}
\bar u_{h^\prime }(\vec{p}_f)\gamma_{\mu} u_{h}(\vec{p}_{i})
A^{\mu }(q )\, ,
&\quad & \bar u_{h^\prime }(\vec{p}_f)\sigma_{\mu\nu} u_h(\vec{p}_i)
F^{\mu \nu}  (q)\, ,\quad q=p_i-p_f\, ,
\end{eqnarray}
etc can be written down and exploited in calculations of 
processes like, say, scattering and bremsstrahlung. 
Moreover, also the construction of higher spins, be they 
matter-, or gauge fields, is straightforward.
{}For example, the massive spin-3/2 spinors in DSR,
be they gravitinos, or matter particles,  
$u^\nu (\vec p\, )$, are constructed as
\begin{eqnarray}
u^\nu _M (\vec p\, )=
\sum_{\mu , h }\left(1\mu \frac{1}{2}h\vert \frac{3}{2}M\right)
\kappa^\prime (\vec p\, )\epsilon_{1\mu }^\nu (\vec 0\, )\otimes
u_h (\vec p\, )\, .
\label{gravitino_DSR}
\end{eqnarray} 
The physics of massive gravitino is important, among others,
in inflationary models of the Universe ~\cite{Moroi}.
{}Finally, the  kinematic anti-screening effect encountered in DSR
is especially intriguing because it has been known so far only in 
non-Abelian theories and attributed to self-interaction of the gauge field.

\vspace{0.2cm}
Work supported by Consejo Nacional de Ciencia y
Tecnolog\'ia (CONACyT, Mexico) under grant number 
C01-39820.


\begin{thebibliography}{99}

\bibitem{9410067}
D.\ J.\  Bird, et al., {  Detection of a cosmic ray with measured 
energy well beyond the expected spectral cutoff due to cosmic 
microwave radiation,} {\it Astrophys.\ J.}  {\bf 441}, 144-150 (1995).

\bibitem{Amelino}
G.\  Amelino-Camelia, { Testable scenario for relativity with 
minimum length\/}, {\it Phys.\ Lett.\ } {\bf B510}, 255-263 (2001).



\bibitem{fb15}  
G.\ Amelino-Camelia, { Relativity  in space-times 
with short distance structure governed by an observer independent  
(Planckian) length scale}, 
{\it Int.\ J.\ Mod.\ Phys.\  } {\bf D11}, 35 (2002).


\bibitem{Giov_Piran} G.\ Amelino-Camelia, Tsvi Piran,
{ Planck-scale deformation of Lorentz symmetry as a solution and
the TeV-$\gamma$ paradoxes }, {\it Phys.\ Rev.\ }
{\bf D64}, 036005 (2001).

\bibitem{Kifune} T.\ Kifune, 
{ Invariance violation extends the comic ray horizon?\/},
{\it Astrophys.\ J.\  } {\bf 518}, L21 (1999).

\bibitem{Ng_vanDam} Y.\ Jack Ng, D.\ -S.\ Lee, M.\ C.\ Oh, 
 H.\ van Dam, { Energy-momentum uncertainties as possible origin of 
threshold anomalies in UHECR and TeV-$\gamma$ events},
{\it Phys.\ Lett.\ } {\bf B507}, 236 (2001).

\bibitem{JV} S.\  Judes, M.\ Visser, { Conservation laws in 
``Doubly Special Relativity''}, 
{\it Phys.\ Rev.\ } {\bf D68}, 045001 (2003). 


\bibitem{fb13} 
J.\ Magueijo, L.\ Smolin, {  Lorentz invariance with an invariant 
energy scale}, 
{\it Phys.\ Rev.\ Lett.\  } {\bf 88}, 190403 (2002).\\
J.\  Magueijo, { New varying speed of light theories}, 
{\it Rept.\ Prog.\ Phys.\ } {\bf 66}, 2025 (2003).


\bibitem{Giovanni_lP} G.\ Amelino-Camelia,
D.\ Benedetti, F.\ D'Andrea, A.\ Procaccini,
{  Comparison of relativity with observer-independent scales of both
velocity and length/mass\/},
{\it Class.\ Quant.\ Grav.\ } {\bf 20}, 5353 (2003).



\bibitem{0203040} 
J.\ Kowalski-Glikman, S.\  Nowak, { Doubly special relativity 
theories as different bases of k-Poincar\'e algebra}, 
{\it Phys.\ Lett.\ } {\bf B539}, 126-132 (2002).


 \bibitem{0304013}
R.\ Lehnert, { Threshold analyses and Lorentz violation},
{\it Phys.\ Rev.\ } {\bf D68}, 085003 (2003).

\bibitem{0207031}
J.\ Rembielinski, A.\ Smolinski, 
{ Unphysical predictions of some doubly special relativity theories }, 
{\it 
Bull.\ Soc.\ Sci.\ Lett.\ Lodz } {\bf 53}, 57-63 (2003).

\bibitem{0308028}
J.\ Christian, { Passage of time in a Planck scale rooted 
local inertial structure},
Int.\ J.\ Mod.\ Phys.\ {\bf D13}, 1037 (2003).





\bibitem{Nilles} H.\ P.\ Nilles, M.\ Peloso, and L.\ Sorbo,
{Nonthermal production of gravitinos and inflatinos}, 
{\it Phys.\ Rev.\ Lett.\ } {\bf  87}, 051302 (2001).

\bibitem{Kallosch} R.\ Kallosh, L.\ Kofman, A.\ Linde, and 
A.\ V.\  Proeyen, 
{ Gravitino production after inflation},
{\it Phys.\ Rev.\  } {\bf D61}, 103503 (2000).

\bibitem{Bolz} M.\ Bolz, W.\ Buchmuller, and M.\ Plumacher,
{ Baryon asymmetry and dark matter},
{\it Phys.\ Lett.\ } {\bf B443}, 209  (1998).

\bibitem{PART} Particle Data Group,
{Review of Particle Physics}, 
{\it Phys.\ Rev.\ } {\bf D66}, 010001-757 (2002).






\bibitem{Hladek} J.\ Hladik, {\it Spinors in Physics\/}
(Springer, N.\ Y.\ , 1999). 




\bibitem{0101009}
 D.\  V.\  Ahluwalia, M.\  Kirchbach, { (1/2,1/2) Representation 
space -An ab initio construct\/},
{\it  Mod.\ Phys.\ Lett.\  }
{\bf A16}, 1377-1383 (2001). 

\bibitem{Valery} V.\ V.\ Dvoeglazov, { Normalization and the 
$m\to 0$ limit of Proca theory\/}, 
{\it Chech.\ J.\ Phys.\ }
{\bf 50}, 119 (2000).

\bibitem{Bety} 
B.\ E.\ Rodriguez-Milla, {\it Baryon excitations as spin-parity
multiplets\/}, University graduate thesis, 
Autonomous University of Zacatecas, M\'exico (2001).


\bibitem{Moroi} T.\ Moroi, Masahiro Yamaguchi, and T.\ Yanagida,
{On the solution to the Polony problem with 0 (10-TeV) gravitino
mass in supergravity\/},
{\it  Phys.\ Lett.\ } {\bf B342}, 105-110 (1995).


\end{thebibliography}
\end{document}